TITLE PAGE

**Depicting deterministic variables within directed acyclic graphs (DAGs): An aid for identifying and interpreting causal effects involving tautological associations, compositional data, and composite variables**

*Laurie Berrie[1] , *Kellyn F. Arnold KF[2], Georgia D. Tomova[2,3,4], Mark S. Gilthorpe[4,5], Peter W.G. Tennant[2,3,4]

[1]*School of GeoSciences, University of Edinburgh, Edinburgh, UK;* [2]*Leeds Institute for Data Analytics, University of Leeds, Leeds, UK;* [3]*School of Medicine, University of Leeds, Leeds, UK;* [4]*The Alan Turing Institute, London, UK;* [5]*Obesity Institute, Leeds Beckett University, Leeds, UK.*

*Lead authors

**Corresponding author**

Peter WG Tennant PhD

*Address*:  Leeds Institute for Data Analytics,

Level 11 Worsley Building,

Clarendon Way,

Leeds,

LS2 9NL,

UK.

*Email*:  P.W.G.Tennant@leeds.ac.uk

*Twitter*:  @PWGTennant

**Acknowledgements**

We thank Sarah Gadd for her contributions to some of the ideas presented in this paper. We also thank Sarah Weiten for reading and commenting on the manuscript.

**Sources of funding**

This work received no specific funding, but during the conduct of this work: LB received funding from the Medical Research Council [MR/K501402/1], KFA received funding from the Economic and Social Research Council [ES/J500215/1] and KFA, GDT, MSG, and PWGT receiving funding from the Alan Turing Institute [EP/N510129/1].

**Conflicts of interest**

PWGT is a director of Causal Thinking Ltd and MSG is a director of Causal Insights Ltd, both of which provide causal inference research and training. Both companies and their directors may benefit from any study that demonstrates the value of causal inference methods.

CONTENTS

**ABSTRACT**

Deterministic variables are variables that are fully explained by one or more parent variables. They commonly arise when a variable has been algebraically constructed from one or more parent variables, as with composite variables, and in compositional data, where the 'whole' variable is determined from its 'parts'.

This article introduces how deterministic variables may be depicted within directed acyclic graphs (DAGs) to help with identifying and interpreting causal effects involving tautological associations, compositional data, and composite variables. We propose a two-step approach in which all variables are initially considered, and an explicit choice is then made whether to focus on the deterministic variable(s) or the determining parents.

Depicting deterministic variables within DAGs bring several benefits. It is easier to identify and avoid misinterpreting tautological associations, i.e., self-fulfilling associations between variables with shared algebraic parent variables. In compositional data, it is easier to understand the consequences of conditioning on the 'whole' variable, and correctly identify total and relative causal effects. For composite variables, it encourages greater consideration of the target estimand and greater scrutiny of the consistency and exchangeability assumptions.

DAGs with deterministic variables are a useful aid for planning and interpreting analyses involving tautological associations, compositional data, and/or composite variables.

**KEY WORDS**

Causal inference, directed acyclic graphs, compositional data, composite variables, deterministic variables, tautological associations

## INTRODUCTION

Causal directed acyclic graphs (DAGs) are increasingly popular aids for identifying and estimating causal effects[1,2] and for recognising and understanding various forms of error, bias, and non-causal associations.[3–10] However, little attention has been given to their utility for understanding analyses involving *deterministic variables*.[11] A deterministic variable is a variable that is fully explained by one or more other variables via a functional non-probabilistic relationship. They are extremely common in health and social science, typically arising in the following three types:

- **Transformed variables** are simple derived variables (e.g., macrosomia) that are functionally created from, and fully explained by, a *single* parent variable (e.g., birthweight). The most common example is **dichotomised variables**, in which a continuous or multinomial parent variable is collapsed into two categories.[12]

- In *compositional data*, also known as comparative data, **'whole' variables** are total variables (e.g., total mass) that contain *two or more* distinct 'part' variables (e.g., fat mass and fat-free mass).[13,14]

- **Composite variables**, also known as compound variables, are complex derived variables (e.g., waist-to-hip ratio) that are functionally created from, and fully explained by, *two or more* parent variables (e.g., waist circumference/hip circumference).[15] Common examples are **ratio variables** (e.g., $X/N$), in which one variable (e.g., $X$) is divided by another (e.g., $N$) and **change score variables** (e.g., $X_1 - X_0$), in which an earlier measure of an evolving variable (e.g., $X_0$) is subtracted from a subsequent measure (e.g., $X_1$)

Because DAGs are primarily used to consider probabilistic relationships,[16] deterministic variables have received limited attention within DAGs. Indeed, DAGs containing deterministic variables have additional statistical implications that make them incompatible with many routine causal identification and discovery algorithms.[17] Depicting deterministic variables within DAGs can, however, be very useful for understanding the unique challenges involved in the analyses and interpretation of deterministic variables.

In the following, we introduce how deterministic variables can be depicted within DAGs and summarise previous approaches to dealing with deterministic variables. We introduce the concepts of *'tautological associations'* and *'composite variable bias'* and discuss how DAGs with deterministic variables can be useful for avoiding misinterpretation. Finally, we consider the challenges with identifying causal effects in compositional data and involving composite variables.

## DEPICTING DETERMINISTIC VARIABLES WITHIN DAGS

A causal DAG is a graphical representation of the hypothesised causal relationships between a set of variables (or 'nodes').[1,2] Any two variables in the graph may be connected by a unidirectional arrow (or 'arc'), which signifies that the first variable (the 'parent' or 'ancestor') exerts a causal effect on the second (the 'child' or 'descendent'). Because a DAG is acyclic, no variable may cause itself. The total causal effect of a specified *exposure* on a specified *outcome* is the combination of all direct and indirect causal effects of the exposure on the outcome. A *confounder* variable causes both the exposure and the outcome, a *mediator* variable is caused by the exposure and causes the outcome, and a *collider* variable is caused by two or more parent variables. When estimating the total causal effect of an exposure on an outcome, conditioning on confounders is encouraged to reduce confounding bias, while conditioning on mediators and colliders is discouraged as it may block part of the total causal effect and/or introduce collider bias.[2,18]

To ensure that deterministic variables are handled appropriately within DAGs they should be distinctively depicted. To achieve this, we follow the convention that any 'child' variable [e.g., $X$] that is fully determined by one or more 'parent' variables [e.g., $pa(X)$] is depicted with a double-outlined node.[19] We also suggest that all arcs entering a deterministic variable should be double-lined, to denote that they are part of a functional, not probabilistic, relationship.[13] Finally, where a child variable and all determining parent variables occur concurrently, we suggest enclosing the family within a dashed-outline box.[13] Examples of this notation are given in **Figure 1**, which depicts a transformed variable (**Figure 1A**), the 'whole' and 'part' variables within compositional data (**Figure 1B**), and a composite variable (**Figure 1C**).

**Figure 1**

Causal directed acyclic graphs using deterministic notation to depict: (A) a derived variable; (B) compositional data; and (C) a composite variable.

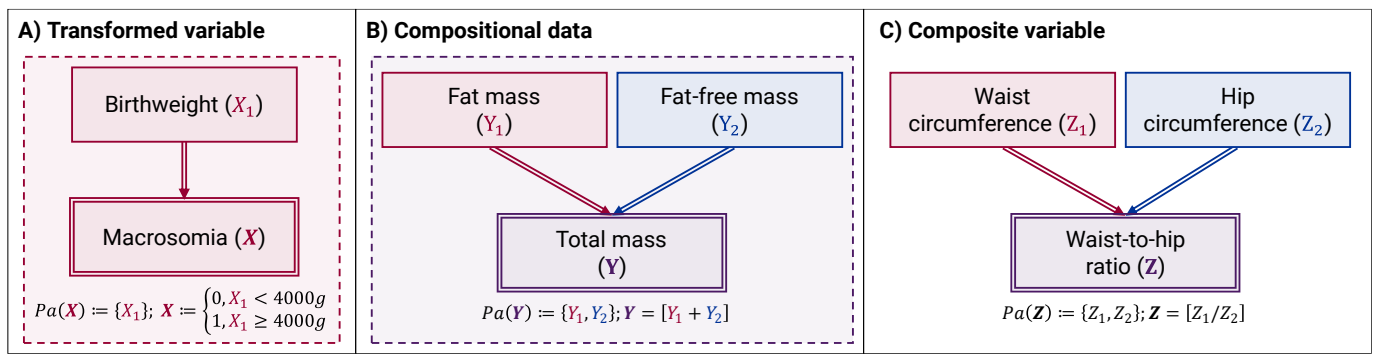

*Fully determined child variables are represented by double-outlined nodes, deterministic relationships are represented by double-lined arcs, and situations where a child variable and its determining parent variables occur simultaneously in time are enclosed within a dashed-outline box. In (A) the transformed variable, macrosomia ($X$), is a binary variable that is fully determined by birthweight ($X_1$). In (B) the 'whole' variable, total mass ($Y$), comprises two 'part' variables, fat mass ($Y_1$) and fat-free mass ($Y_2$) and can therefore be fully determined by summing both parent components. In (C) the composite variable, waist-to-hip ratio (Z), is fully determined by dividing the waist circumference [$Z_1$] by the hip circumference [$Z_2$].*

## ALGORITHMIC APPROACHES AND THE BENEFIT OF DAGS

For many years, deterministic variables were not strictly compatible with DAGs because deterministic variables bring additional statistical dependencies. This was resolved with the introduction of the D-separation criterion (note the uppercase 'D'), which extends the familiar d-separation criterion to accommodate the behaviour of deterministic variables.[19] Despite this, most causal modelling and discovery algorithms are not natively compatible with deterministic variables.[17] Such variables are hence usually treated as nuisance nodes that need to be identified and removed.[19] Shachter's 'Deterministic Node Reduction' algorithm achieves this by identifying all deterministic variables within a graphical network and transferring the incoming and outgoing probabilistic arcs to their parent nodes to create 'barren nodes' with no onward arcs that may be removed without losing information about the relationship between the remaining variables.[11]

Deterministic Node Reduction was however developed before DAGs evolved to encode *causal* information.[16] In a *causal* DAG, the inclusion of a transformed or composite variable offers a way to encode information about the *nature* of a causal relationship. For example, Attia *et al.* propose using deterministic multiplicative nodes, such as 'A×B', to depict causal interactions and/or effect modification.[20] Simply identifying and discarding deterministic variables does not therefore offer a universal solution to the handling of such variables.

We advocate an alternative two-step approach to considering deterministic variables. First, a 'full' DAG is drawn that includes all deterministic variables and all determining parents. Next, an explicit choice is made whether to focus on the deterministic variable(s) or the determining parents. This approach ensures that the context and analysis are given sufficiently thorough attention, reducing the risk of misinterpretation.

**UNDERSTANDING TAUTOLOGICAL ASSOCIATIONS**

Perhaps the most straightforward benefit of depicting deterministic variables within DAGs is the ability to identify and avoid misinterpreting tautological associations. We define a tautological association as a 'self-fulfilling' association that arises when a variable is analysed in relation to itself, a part thereof, or a 'sibling' variable with at least one shared parent component.

The problem of tautological associations was first identified by Karl Pearson in 1897.[21] Assuming faithfulness,[1] Pearson warned that two ratio variables with a shared denominator parent variable (e.g., $X/N$ and $Y/N$) will share a '*spurious (organic) correlation*' even if the numerators (e.g., $X$, $Y$) are unrelated.[21] Using deterministic notation, this can be depicted and understood using a DAG that contains the three parent variables (i.e., $X$, $Y$, and $N$) and the two child variables ($X/N$ and $Y/N$) (**Figure 2A**).

Another common example is the tautological association between a 'change score' variable (e.g., $X_1 - X_0$) and the baseline parent variable (e.g., $X_0$) (**Figure 2B**). First explained by Oldham in 1962,[22] the resulting negative correlation between the change score and the baseline parent variable has caused considerable confusion.[22–24] Other examples can be found throughout the literature, often with the term '*mathematic(al) coupling*' coined by Archie in 1981,[23] although most examples probably occur in applied analyses with no awareness of the phenomenon.

In both statistical and causal terms, tautological associations are neither erroneous nor biased.[25] The expected association between two ratio variables, for example, is an accurate reflection of their common denominator variable.[25] Tautological associations are however commonly misinterpreted when the underlying tautology is not recognised, and the resulting associations are misattributed to other mechanisms. Such misinterpretations are an example of *composite variable bias*, which we define as the collection of biases and fallacies that complicate the identification, analysis, and interpretation of causal effects involving composite variables.

The misinterpretation of tautological associations is probably more common for composite variables with many parent variables since the deterministic origins become easier to overlook. Nevertheless, there are examples of tautological associations between transformed descendants of the same parent variable being overlooked when analyses are conducted at aggregate level (**Figure 2C**).[26] By placing all parent and child variables within a DAG, we believe such mistakes become less likely.

**Figure 2**

Causal directed acyclic graphs of three tautological associations: (A) between two composite ratio variables with a common denominator parent variable; (B) between a change-score variable and its baseline parent variable; and (C) between two aggregate variables.

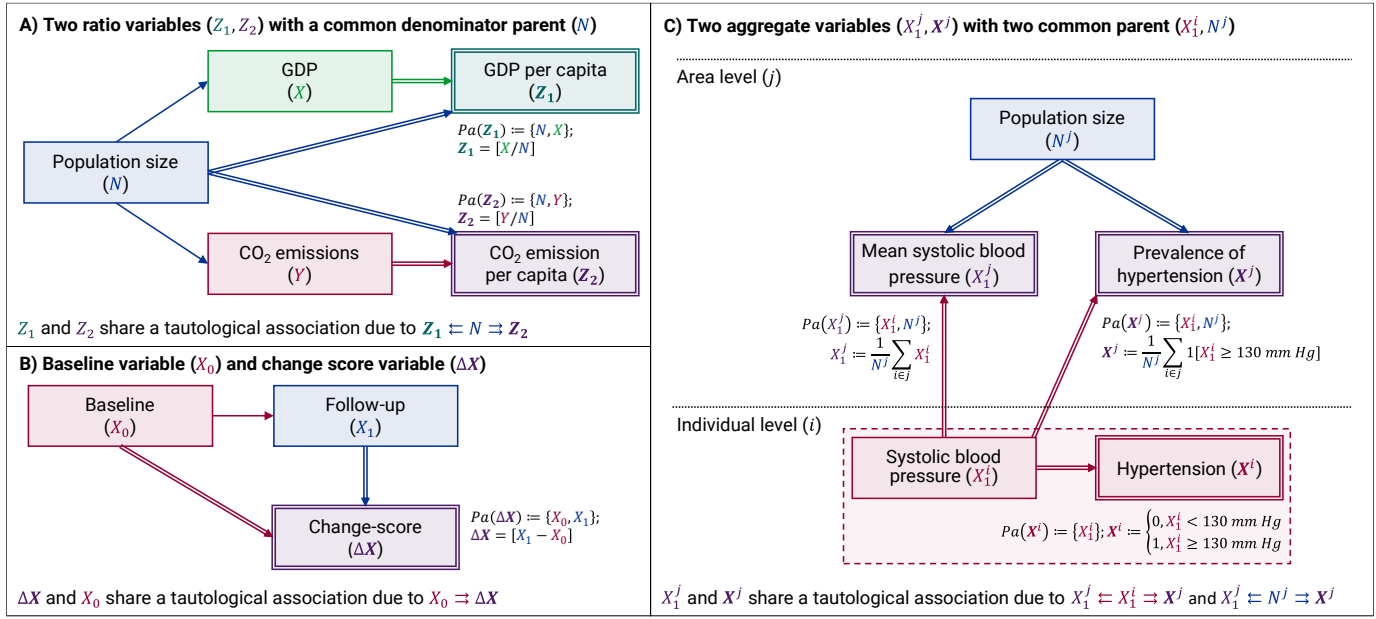

In (A) the observed variables GDP ($X$) and $CO_2$ emissions ($Y$) are both caused by population size ($N$), making $N$ a confounder for the apparent relationship between $X$ and $Y$. Two composite variables have also been created by diving both $X$ and $Y$ by $N$ to create GDP per capita ($Z_1$) and $CO_2$ emission per capita ($Z_2$) respectively. Since both $Z_1$ and $Z_2$ share the same parent variable ($N$), they will share a tautological association. In (B) the baseline measurement of a repeated measure variable ($X_0$) causes the follow-up measurement ($X_1$), from which a composite change score variable ($\Delta X$) has been created by subtracting $X_0$ from $X_1$. Since $X_0$ is a deterministic parent of $\Delta X$, they share a tautological association. In (C), where subscript $i$ denotes individual-level and $j$ denotes area-level, a transformed variable, hypertension ($X^i$), is a binary individual-level variable that is fully determined by the continuous individual-level variable systolic blood pressure ($X_1^i$). Area-level mean systolic blood pressure ($X_1^j$) and area-level prevalence of hypertension ($X^j$) are determined at the aggregate level from $X_1^i$ and the area-level population ($N^j$). Since both $X_1^j$ and $X^j$ share the same two parent variables ($X_1^i, N^j$), they share a tautological association.

## DEPICTING AND CONSIDERING COMPOSITIONAL DATA

The benefits of depicting deterministic variables within DAGs increases with the complexity of the variables and/or relationships being considered. Compositional data is a common form of complex data structure that naturally contains deterministic relationships because the 'part' variables sum to one or more 'whole' variables (**Figure 1B**).[13,14] This makes compositional data notoriously challenging to analyse and interpret correctly.[13,14,27,28] Pearson's warning on the use of ratio variables was allegedly motivated by witnessing biologists dividing bone measurement variables (e.g., femur length) by length measurement variables (e.g., leg length).[21] Since then, the area of Compositional Data Analysis (CoDA) has emerged to develop alternative analytical strategies.[14,29] We focus on the practice and insights of depicting compositional data within DAGs.[13,27,28]

**Simplifying features of compositional data**

There are two distinct and important features of compositional data that reduce the potential for analytical and interpretational issues when compared with composite variables. First, the 'whole' variable can be directly observed. Indeed, whether a variable is a 'whole' or 'part' is usually a matter of perspective or convenience rather than external structure. All variables can potentially be divided into further parts or summed to a greater whole. The choice of whether to focus on the 'whole' or the 'parts' is therefore usually a trade-off between the competing benefits of aggregation and subdivision.

The second key feature of compositional data is that the 'whole' variable and all 'part' variables 'crystallise' at the same time, meaning they all belong in the same place within the DAG. This avoids many of the more serious issues affecting composite variables discussed below.

**Choosing the target estimand**

Analyses of compositional data generally consider two types of estimands: total causal effects and relative causal effects. Total causal effects in compositional data represent the effect of increasing the 'whole' variable either by intervening on the 'whole' directly or through one or more specified 'part' exposures; these are sometimes known as 'additive effects'.[13,27,28,30] Relative causal effects in compositional data represent the joint effect of increasing a specified 'part' exposure while simultaneously decreasing one or more substituting 'parts' to keep the 'whole' fixed; these are sometimes known as 'substitution effects'.[13,27,28,30] Different analytical strategies are required to estimate these two effects and misinterpretations occur when the wrong strategy is used inadvertently.[13,27,28] That said, in some contexts (e.g., time spent sleeping, physically active, and sedentary) the 'whole' may be structurally fixed (e.g., because there are only 24 hours in the day), meaning that only relative causal effects can be obtained.[13]

Since compositional data occur at the same time, the 'whole' and 'part' variables may be drawn in multiple ways, but it is intuitive to consider the 'whole' as being fully determined by the 'parts' (**Figure 1B**).[13] Drawn like this, the 'whole' can be usefully interpreted as a *collider* for the 'parts', and it is clear that conditioning on the 'whole' will introduce a dependency between the 'parts'.[13] The total causal effect of a specific 'part' cannot therefore be estimated when conditioning on the 'whole'.[13]

To illustrate, we consider the total causal effect of carbohydrate consumption on the risk of diabetes, where the consumption of carbohydrates, proteins, and fats determines the total energy intake (**Figure 3A**). In nutrition research, it is common to evaluate the effects of one or more specific dietary component(s) on subsequent health outcomes while conditioning on total energy intake as a proxy "confounder" for the diet.[31] However, when drawn as suggested, it is apparent that conditioning on total energy intake would introduce a dependency between carbohydrate consumption and the other macronutrient variables, creating a relative effect (**Figure 3B**). Exactly which relative effect is estimated will depend on whether additional adjustments are made for any of the other macronutrients (**Figure 3C**).

Relative effect estimates are sometimes also sought by analysing ratio variables, created by dividing each 'part' variable by the 'whole'.[27,32] Unfortunately, such analyses are extremely susceptible to composite variable bias, because the resulting ratio variables conflate the effects of their numerator 'part' parent variable with the inverse of their denominator 'whole' parent variable (**Figure 3D**).[33] We discuss this further when considering composite variables below.

**Figure 3**

Causal directed acyclic graphs (DAGs) examining the identification of causal effects in compositional data, here represented by total energy intake.

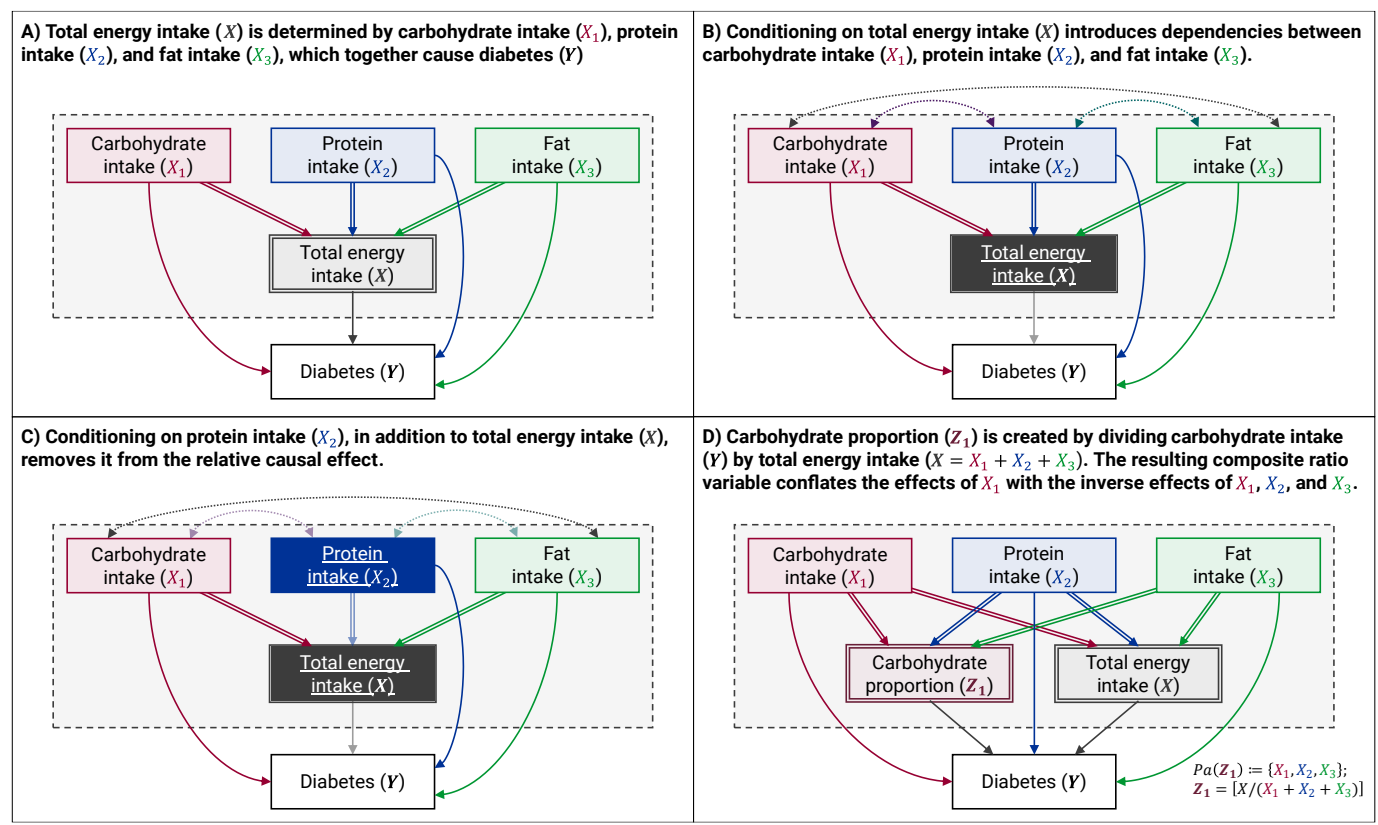

*The 'full' DAG for the scenario is depicted in (A), where the 'whole' variable (total energy intake, $X$) is fully determined by intake from three 'part' variables (carbohydrate intake, $X_1$, protein intake, $X_2$, and fat intake, $X_3$), which together cause diabetes ($Y$). In (B) the 'whole' has been conditioned on, inducing conditional dependencies between the unconditioned 'part' variables. The causal effect of any 'part' variable (e.g., $X_1$) on $Y$ would thus be relative to the other unconditioned 'part' variables (e.g., $X_2$, and $X_3$). In (C) one of the other 'part' variables ($X_2$) has been conditioned on in addition to the whole, which removes it from the relative causal effect; the effect of $X_1$ on $Y$ will thus be relative to $X_3$ only. In (D) a composite ratio variable, carbohydrate proportion ($Z_1$), is introduced by dividing one of the parts, carbohydrate intake ($X_1$), by the 'whole' variable, total energy intake ($X = X_1 + X_2 + X_3$). The resultant composite ratio variable ($Z_1$) conflates the effects of the numerator parent variable ($X_1$) with the inverse of the effects of the denominator parent variables ($X_1, X_2, X_3$).*

## Identifying and estimating causal effects in compositional data

This simple three-nutrient example demonstrates how retaining both the 'whole' and 'part' variables within a DAG can help with understanding compositional data. The optimal analytical strategy then depends on whether the investigator is interested in the total causal effect of a particular 'part', a relative causal effect of a particular 'part', or the summary effect of the 'whole'. There are, however, additional caveats.

First, the exchangeability assumption requires that the units of analysis have a conditionally equivalent probability of the outcome at the time of exposure, i.e., there must be no unobserved or residual confounding for our exposure-outcome relationship of interest. In compositional data, confounding can arise from common causes of the 'parts', even if these causes have no residual effect on the outcome, because each part is itself likely to cause the outcome. In our three-nutrient example, such common causes might include the specific foods consumed. Ideally, such common causes should be directly measured and conditioned on, but this is not always possible. Instead, when interested in the total causal effect of a particular 'part', confounding by common causes may be reduced by conditioning on other 'parts' to block the confounding paths downstream. Where each 'part' has a unique effect and variance, this requires measuring and conditioning on every part variable. In practice, aggregated 'part' variables are often used, such as 'remaining energy intake' (i.e., energy from all parts *except* the exposure), but this may introduce residual confounding wherever the causal effect of each 'part' differs from the average effect of the aggregate variable.[27,28]

The same problem can affect estimates of relative causal effects when adjusting for the 'whole'. The most accurate way to estimate a relative causal effect for a particular component when adjusting for the 'whole' is therefore to condition on all individual components that are not involved in the substitution. Relative causal effects can also be obtained by estimating the total causal effects of 'all components' and directly comparing the coefficients for the exposure and substituting component of interest.[27,28]

In some situations, the average effect of increasing the 'whole' may be of more interest than the individual 'parts' specifically. Here, it may be reasonable to discard the parent variables from the DAG and treat the 'whole' variable as the exposure. However, this increases the chance of violating the consistency assumption, which requires that there are '*no two "flavors" or versions of treatment*',[34] since the same value of the 'whole' can be obtained from many different combinations of the 'parts'. If each 'part' has different causal effects on the outcome and/or different variances, then the summary effect of the 'whole' will not be the mean-weighted average effect but will be distorted towards those 'parts' with the largest variances.[15,27,28]

In theory, measuring and modelling all components offers the ideal approach to compositional data analyses. In practice, the benefits of achieving greater consistency need balancing against the demands of modelling ever more variables. The probability of experiencing positivity violations and finite sample bias increases with the number of components.[35] The choice of whether to focus on the 'whole' or the 'parts' will therefore involve balancing the desired degree of consistency with the quality and availability of the data.

## DEPICTING AND CONSIDERING COMPOSITE VARIABLES

Within a DAG, a composite variable appears similar to a compositional 'whole' variable, with two or more parents causing a fully determined child (**Figure 1C**). There are however some important features of composite variables that make them particularly prone to composite variable bias. First, composite variables cannot be directly measured; they can only be known once the parents are themselves known.[36] Second, the parent variables may be subject to a range of functional transformations besides addition, including subtraction, division, and exponentiation.[15] Finally, the parent components may not crystallise concurrently, giving them different temporal positions within the DAG. It is therefore extremely important to consider the nature and purpose of every composite variable that is used in causal inference.

### Choosing the target estimand

Composite variables are commonly constructed for one of two reasons:

- To *summarise* several interrelated variables (e.g., deprivation index) into a single variable, either to capture a latent concept (e.g., socioeconomic circumstances), define a multifactorial state (e.g., metabolic syndrome), or provide a global summary (e.g., Disease Activity Scores)

- To *standardise* one or more variables against one or more other variables (e.g., GDP per capita), either to account for another variable (e.g., population size), or rescale to a common unit (e.g., percentage change).

Whether a composite variable has been constructed to summarise or to standardise has immediate implications for its analysis and/or interpretation. The creation of a summary variable implies an interest in estimating the average effect of, or on, the parent variables. Conversely, composite variables that seek to standardise imply an interest in one or more parent variables while controlling for one or more 'nuisance' components. Ratio variables and change score variables, for example, both seek to isolate one parent from another, using division and subtraction, respectively.[21,24,25] Unfortunately, such approaches only transform, rather than remove, the nuisance components.[21,24,25,27] Causal analyses of ratio variables and change score variables are hence particularly prone to composite variable bias.[21,24,25,27] For most standardised composite variables, it is likely that the true target of interest is one or more target parent variables *conditional* on one or more nuisance parent variables. Rather than attempting an algebraic solution, such conditioning should be attempted using an appropriate approach, such as covariate adjustment within a linear regression model.

To illustrate, we consider the causal effect of body mass index (BMI = weight/height-squared) at age 50 years on the risk of cardiovascular disease (**Figure 4A**). In probabilistic terms, BMI contributes no information beyond what is captured by weight and height.[37] Deterministic Node Reduction would hence reduce BMI to a barren node that may be removed without losing information about the relationship between the remaining variables (**Figure 4B**). This explains previous assertions that '*no causal knowledge is gained by estimating a nonexistent effect of body mass index*'.[37] However, BMI may still have some utility if it provides a useful parametric summary of the *nature* of the causal effects of weight and height.

Since BMI is constructed by dividing weight by height-squared, it seems reasonable to assume it was conceived to standardise weight by height. That said, inventor Adolphe Quetelet (1832) offers no specific motivation beyond reporting that, '*weight increases approximately with the square of the height*'.[38] Furthermore, Keys *et al*. (1972), who transformed the name and prominence of the index, only appear interested in finding the best proxy measure of skinfold thickness.[39] Whether BMI is intended purely as a measure of weight standardised for height or a summary proxy for adiposity cannot therefore be known from history or algorithm, but the two perspectives carry different implications. If BMI is hypothesised as a valuable joint summary of weight and 1/height-squared then focussing on the composite measure may be reasonable, notwithstanding the issues discussed below (**Figure 4C**). Alternatively, if BMI is simply a measure of weight standardised by height, then the appropriate target would be weight *conditional on height* (**Figure 4B**). Since the two approaches likely provide different results, determining the true estimand of interest is clearly extremely important.

**Figure 4**

Causal directed acyclic graphs (DAGs) considering the causal effect of a composite variable (body mass index, BMI) on an outcome (cardiovascular disease, CVD).

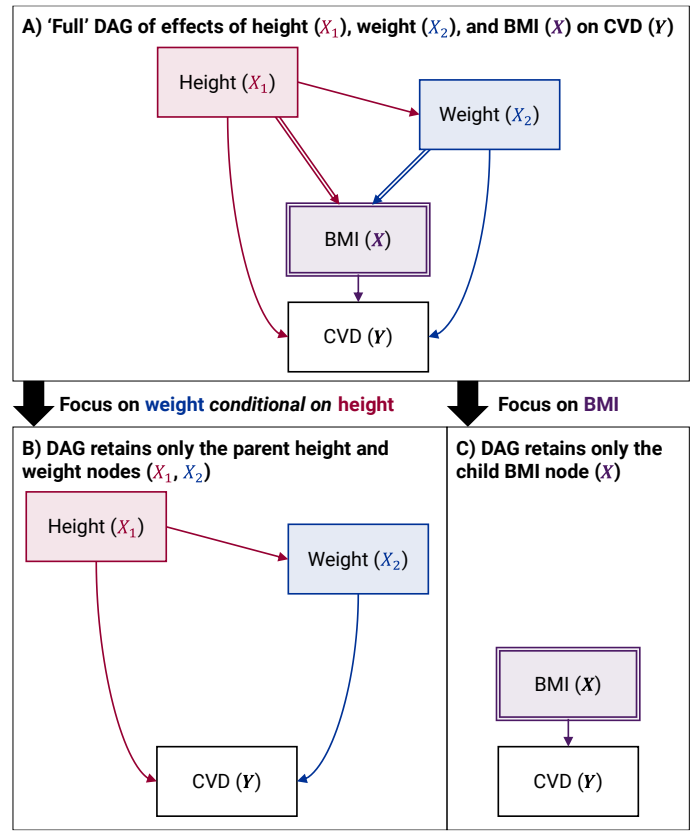

*The 'full' DAG in (A) shows BMI as a fully determined child of height and weight. An explicit choice can be made, based on the target estimand of interest, to either retain and focus on the parent nodes, as shown in (B), or to retain and focus on the child node, as shown in (C).*

**Identifying and estimating causal effects involving composite variables**

Regardless of their potential utility, most composite variables are likely to experience issues with satisfying the consistency and exchangeability assumptions. As with compositional 'whole' variables, composite variables have an inherent risk of consistency violations because the same value can be obtained from many different combinations of the parents. However, since composite variables are typically made from a more heterogeneous mix of parent variables than the compositional 'whole', the impact of these violations may be more severe. When only the summary effect of the composite is available, the individual parent effects are lost and it becomes impossible to know which parent variables are responsible and to what extent.[40] Furthermore, the summary effect of the composite is likely overly influenced by those variables with the largest variation within the sample,[15] leading to sample-specific effects that may transport poorly.

These problems can all be recognised in our BMI example. The same value of BMI can be achieved by infinite combinations of weight and height. Where an effect of BMI is observed, it is impossible to know whether it is due to weight, height, or a combination of both. This alone makes it difficult to translate an observed effect into a practical intervention. However, since height has been inversed and squared, most of the variance of BMI in an adult population will likely be driven by the variation in height. Any small causal relationship between height and either the exposure or outcome is therefore likely to dominate the summary relationship with BMI.

Regardless, the exchangeability assumption is likely the greatest barrier to identifying the causal effect of, or on, a composite variable. In theory, robustly identifying the causal effect of, or on, a composite variable requires that all confounding paths are closed for all parent variables. Unfortunately, when working with the composite parent variable alone, the unique paths to and from each parent variable become conflated. Attempts to block confounding paths may therefore experience residual confounding, since only the diluted summary effect is modelled. More concerningly, if the parent variables crystallise at different moments in time, it is possible they will have different relationships with the supposed confounders. Indeed, it is possible that a confounder for one parent variable may be a mediator for another, leaving no means to appropriately adjust for confounding without also blocking part of the true causal effect.

To illustrate, we return to our BMI example by introducing another variable $C$, which is thought to confound the relationship between BMI at age 50 and the risk of cardiovascular disease (**Figure 5**). When included in a parentless DAG, $C$ might appear to be an unremarkable confounder for the BMI variable, which has been (naively) depicted as occurring at a single point in time (**Figure 5A**). However, the parent variables, adult height and adult weight, do not crystallise at the same time. Typically, height is fixed in early adulthood while weight is a time-varying variable that is likely to crystallise closer to the time of measurement. Depending on when $C$ occurs, it might therefore have a very different relationship with the two parent components of BMI. For example, if $C$ crystallised in infancy (e.g. duration of breastfeeding), then it would likely cause both weight and height in adulthood and be an uncomplicated confounder for the effect of BMI on cardiovascular disease (**Figure 5B**). However, if $C$ crystallised during adolescence (e.g. sugar consumption at secondary school), then it would likely have a stronger effect on current adult weight than current adult height (**Figure 5C**). Finally, if $C$ crystallised in adulthood (e.g. marital status), then it is plausible that while it might cause weight,[41] it might be *caused by* height (**Figure 5D**).[42] Although each of these specific examples can be debated, the 'true' $C$ is likely to represent multiple variables, each of which may have different relationships with the individual components of BMI. The time between the crystallisation of adult height and adult weight therefore makes the causal effect of BMI at age 50 years on the risk of cardiovascular disease impossible to identify and estimate.

**Figure 5**

Causal directed acyclic graphs (DAGs) examining the challenges of identifying a causal effect for a composite variable (body mass index, BMI) when the parent variables are separated in time.

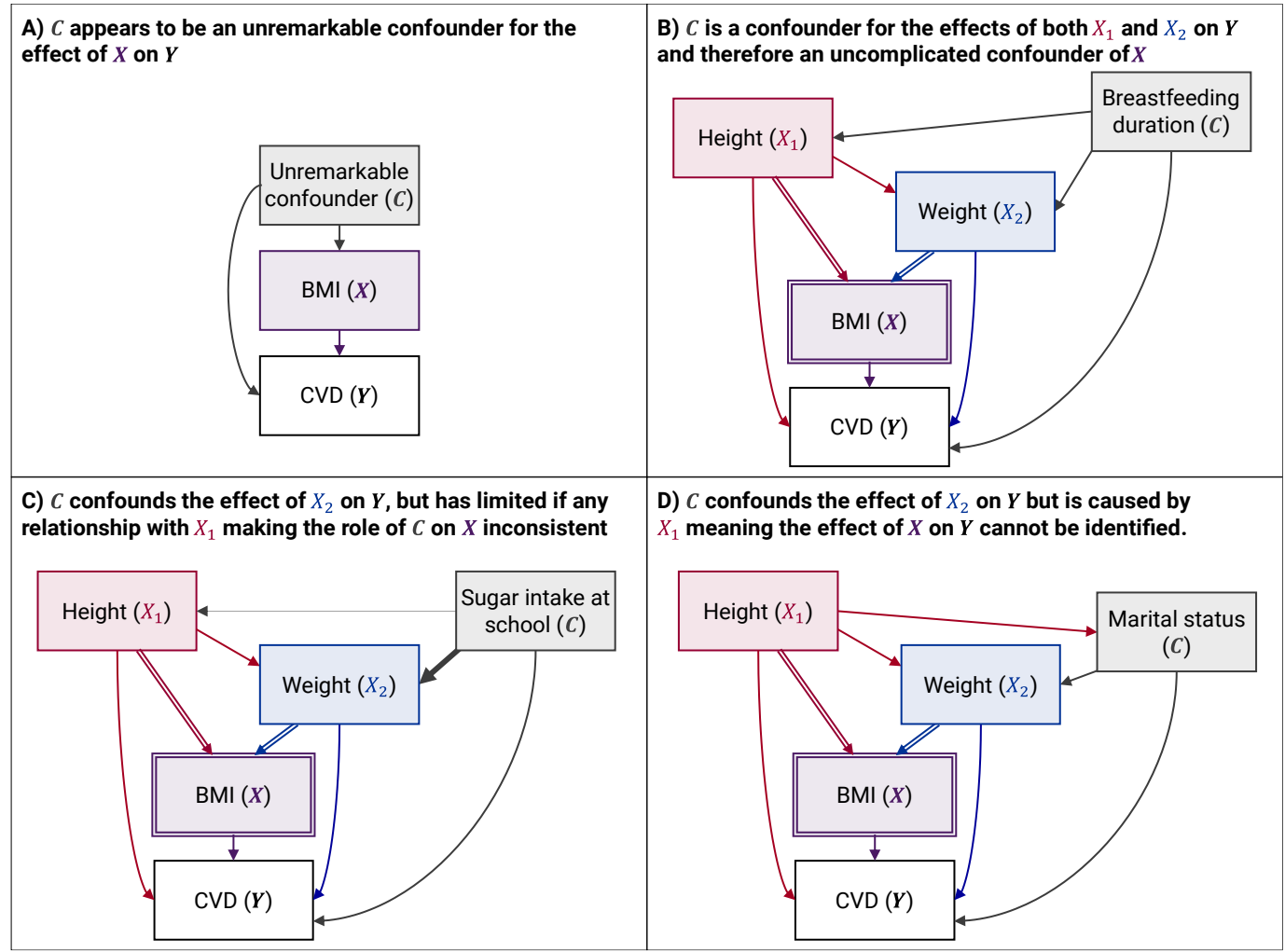

When a composite variable, such as body mass index (BMI, **X**), is included in a DAG without including the parents, the distinct relationship with each component may be overlooked. In (A) the variable **C** appears to be an unremarkable confounder for the effect of BMI on cardiovascular disease (CVD, **Y**), but **C** may have very different relationships with the parent variables height ($X_1$) and weight ($X_2$), which do not crystallise at the same point in time. In (B) the variable **C** (e.g., breastfeeding duration) crystallises long before both parent components have crystallised and has a similar effect on both, making it an uncomplicated confounder of BMI. In (C) the variable **C** (e.g., sugar consumption at secondary school) crystallises alongside both components; but has a much stronger effect on weight, making it an inconsistent confounder of BMI. Finally, in (D) the variable **C** (e.g., marital status) crystallises after one component (height, $X_1$) but before the other (weight, $X_2$), making it a confounder for the effect of weight on CVD, but a mediator for the effect of height on CVD. In this scenario, the summary effect of the composite variable (BMI, **X**) on the outcome cannot be identified.

# CONCLUSION

Deterministic variables are ubiquitous in health and social science research due to the widespread use of transformed and composite variables and the common occurrence of compositional data. Unfortunately, despite repeated warnings over many decades,[21–25,43] the analytic and interpretational challenges of such variables remain largely underappreciated. With appropriate care and notation, we believe that DAGs can provide a novel and effective means to transform our recognition and understanding of these issues. We therefore encourage researchers to consider including deterministic variables in their DAGs when they are planning and/or interpreting analyses involving transformed variables, compositional data, and/or composite variables.